\documentclass[a4paper]{jpconf}
\usepackage{graphicx}

\usepackage{url}
\usepackage{hyperref}
\usepackage{color}
\usepackage{units}
\usepackage{booktabs} 

\begin{document}
\definecolor{brown}{RGB}{139,69,19}

\newcommand{\nuc}[2]{$^{#2}\rm #1$}

\newcommand{\bb}[1]{$\rm #1\nu \beta \beta$}
\newcommand{\bbm}[1]{$\rm #1\nu \beta^- \beta^-$}
\newcommand{\bbp}[1]{$\rm #1\nu \beta^+ \beta^+$}
\newcommand{\bbe}[1]{$\rm #1\nu \rm ECEC$}
\newcommand{\bbep}[1]{$\rm #1\nu \rm EC \beta^+$}

\newcommand{\largeGERDA}{{LArGe}}
\newcommand{\PI}{\mbox{\textsc{Phase\,I}}}
\newcommand{\PIa}{\mbox{\textsc{Phase\,Ia}}}
\newcommand{\PIb}{\mbox{\textsc{Phase\,Ib}}}
\newcommand{\PIc}{\mbox{\textsc{Phase\,Ic}}}
\newcommand{\PII}{\mbox{\textsc{Phase\,II}}}

\newcommand{\aoe}{$A/E$}

\newcommand{\order}[1]{\mbox{$\mathcal{O}$(#1)}}

\newcommand{\pic}[5]{
       \begin{figure}[ht]
       \begin{center}
       \includegraphics[width=#2\textwidth, keepaspectratio, #3]{#1}
       \end{center}
       \caption{#5}
       \label{#4}
       \end{figure}
}

\newcommand{\apic}[5]{
       \begin{figure}[H]
       \begin{center}
       \includegraphics[width=#2\textwidth, keepaspectratio, #3]{#1}
       \end{center}
       \caption{#5}
       \label{#4}
       \end{figure}
}

\newcommand{\sapic}[5]{
       \begin{figure}[P]
       \begin{center}
       \includegraphics[width=#2\textwidth, keepaspectratio, #3]{#1}
       \end{center}
       \caption{#5}
       \label{#4}
       \end{figure}
}

\newcommand{\picwrap}[9]{
       \begin{wrapfigure}{#5}{#6}
       \vspace{#7}
       \begin{center}
       \includegraphics[width=#2\textwidth, keepaspectratio, #3]{#1}
       \end{center}
       \caption{#9}
       \label{#4}
       \vspace{#8}
       \end{wrapfigure}
}

\newcommand{\baseT}[2]{\mbox{$#1\cdot10^{#2}$}}
\newcommand{\baseTsolo}[1]{$10^{#1}$}
\newcommand{\THL}{$T_{\nicefrac{1}{2}}$}

\newcommand{\UBI}{$\rm cts/(kg \cdot yr \cdot keV)$}

\newcommand{\Uflux}{$\rm m^{-2} s^{-1}$}
\newcommand{\Ucpd}{$\rm cts/(kg \cdot d)$}
\newcommand{\Uexpo}{$\rm kg \cdot d$}

\newcommand{\Qbb}{$Q_{\beta\beta}$}

\newcommand{\validate}{\textcolor{blue}{\textit{(validate!!!)}}}

\newcommand{\improve}{\textcolor{blue}{\textit{(improve!!!)}}}

\newcommand{\missing}{\textcolor{red}{\textbf{...!!!...} }}

\newcommand{\quanta}{\textcolor{red}{\textit{(quantitativ?) }}}

\newcommand{\misscite}{\textcolor{red}{[citation!!!]}}

\newcommand{\missref}{\textcolor{red}{[reference!!!]}\ }

\newcommand{\PC}{$N_{\rm peak}$}
\newcommand{\BIC}{$N_{\rm BI}$}
\newcommand{\PAPR}{$R_{\rm p/>p}$}

\newcommand{\PCR}{$R_{\rm peak}$}


\newcommand{\gline}{$\gamma$-line}
\newcommand{\glines}{$\gamma$-lines}

\newcommand{\gray}{$\gamma$-ray}
\newcommand{\grays}{$\gamma$-rays}

\newcommand{\bray}{$\beta$-ray}
\newcommand{\brays}{$\beta$-rays}

\newcommand{\betas}{$\beta$'s}


\newcommand{\tab}{\textcolor{brown}{Tab.~}}
\newcommand{\eq}{\textcolor{brown}{Eq.~}}
\newcommand{\fig}{\textcolor{brown}{Fig.~}}
\renewcommand{\sec}{\textcolor{brown}{Sec.~}}
\newcommand{\chap}{\textcolor{brown}{Chap.~}}

 \newcommand{\fn}{\iffalse \fi} 
 \newcommand{\tx}{\iffalse \fi} 
 \newcommand{\txe}{\iffalse \fi} 
 \newcommand{\sr}{\iffalse \fi} 

\title{Dark Matter Search with the Nuclear Isomer $^{\bf 180m}$Ta}

\author{Bjoern Lehnert}

\address{Nuclear Science Division, Lawrence Berkeley National Laboratory, Berkeley, CA 94720, U.S.A.}

\ead{bjoernlehnert@lbl.gov}

\begin{abstract}
There is compelling cosmological and astrophysical evidence of dark matter comprising 27\% of the energy budget of the Universe. However, dark matter has never been observed in direct detection experiments. The long-time favorite model of Weakly Interacting Massive Particles saw a large experimental effort with steady progress over recent decades. Since also these large-scale searches remain unsuccessful to date, it is compelling to look at more exotic dark matter models which can be constrained with new approaches and much less scientific resources. Using nuclear isomers is one of these approaches.

\nuc{Ta}{180m} is the rarest known isotope with the longest-lived meta-stable state whose partial half-life limits are on the order of $10^{14016}$~yr. We investigate how strongly interacting dark matter and inelastic dark dark matter collides with \nuc{Ta}{180m}, leading to its de-excitation.
The energy stored in the meta-stable state is released in the transition, which becomes the signature for thermalized dark matter in a well-shielded underground experiment.

We report on a direct detection experiment searching for these dark-matter-induced decay signatures which has further constrained the open parameter space. 
We also propose an indirect geochemical experiment to search for decay products of \nuc{Ta}{180m} in tantalum minerals accumulated over 1 billion years.\end{abstract}

\section{Introduction}

Many cosmological observations suggest dark matter (DM) as the dominant form of matter in the Universe making up around 27\% of its energy density. 
There is a wide range of possibilities for the nature of DM from wave-like Axions at the $\mu$eV scale, via particle-like WIMPs at the GeV and TeV scale, and very heavy objects up to the \baseTsolo{13} TeV scale.    
Among those hypotheses, Weakly Interactive Massive Particles (WIMPs) were and are one of the favorite candidates with a natural fit to the primordial density requirements and typical interaction strength of the weak force.
The search for DM is an active field of research with many experiments exploring different hypotheses. 
Since direct detection experiments remain unsuccessful to date, the community is increasingly starting to explore more exotic hypotheses which could have evaded previous searches. This could be the case if the cosmologically observed DM density is comprised of a variety of different forms or if DM is made out of composite particles in bound states.   
 
One of these exotic hypotheses is strongly interacting WIMP DM. 
If the interaction strength with normal matter is too strong, the DM particles would thermalize in the overburden of underground laboratories and loose their galactic velocity distribution. Consequently, they do not have enough kinetic energy to create a detectable nuclear recoil in particle detectors underground. 
On the other hand, such strongly interacting DM would be visible by surface detectors or ballon and rocket experiments. 
Those experiment exclude most of the larger cross sections if all the DM were strongly interacting. However, they are not sensitive enough to exclude this hypothesis if only a fraction of the DM density is of this form (see e.g.\ \cite{Digman:2019}).
Nuclear isomers can help to increase the sensitivity for these searches, providing their stored nuclear excitation energy as mini DM accelerators when interactions occur \cite{Pospelov:2019}.     
 
 \begin{figure}[h]
\centering
\includegraphics[width=0.63\textwidth]{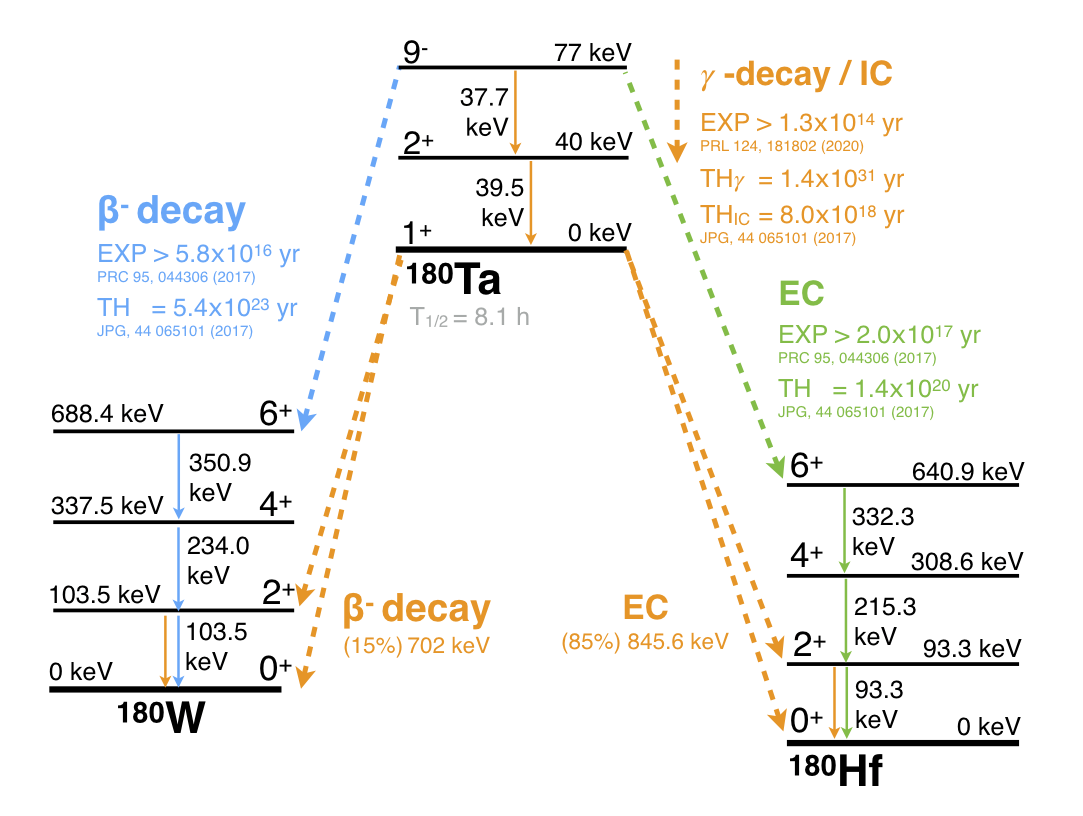}%
\caption{\label{pic:DecayScheme} Decay scheme of \nuc{Ta}{180m} showing the $\beta^-$ decay branch, the electron capture branch, and the internal conversion / $\gamma$ decay branch in different colors. Also shown are the experimental signatures for \gray\ spectroscopy as well as the existing experimental constraints and theoretical predictions.  }
\end{figure}  

\nuc{Ta}{\rm 180m} is a long lived meta stable state and the rarest quasi stable isotope known. 
\nuc{Ta}{\rm 180} exclusively occurs in the 77~keV excited $9^-$ state and is by far the the longest lived isomer with a half-live exceeding \baseTsolo{14}~yr.
Its decay via internal conversion (IC) or $\gamma$-decay into the lower $2^+$ state or the $1^+$ ground state is suppressed, requiring the change of parity and a spin change of 7.  
In addition, its decay is possible through $\beta$-decay into \nuc{W}{180} and electron capture (EC) into \nuc{Hf}{180}. But these decay modes are also strongly suppressed requiring 3-fold non-unique transitions into the excited $6^+$ states of the daughters.  
All decay modes are shown in \fig \ref{pic:DecayScheme} including current half-life constraints from \cite{Lehnert:2017} and \cite{Lehnert:2020} and theoretical predictions from \cite{Ejiri:2017js}.

The de-excitation of \nuc{Ta}{180m} via DM has been recently suggested in \cite{Pospelov:2019,Lehnert:2020}. The large mass of DM can absorb the angular momentum change in the scatter populating the $2^+$ or $1^+$ states of \nuc{Ta}{180}. This is followed by the consecutive decay into \nuc{W}{180} and \nuc{Hf}{180} from the \nuc{Ta}{180} ground state. 
Notably, the decay from \nuc{Ta}{180m} and \nuc{Ta}{180} have different branching rations populating \nuc{W}{180} and \nuc{Hf}{180} differently. The experimental signatures for the \nuc{Ta}{180m} decay include easily detectable high energy \grays\ up to 351~keV, whereas the \nuc{Ta}{180} decay only produces low energy \grays\ of 103.5 and 93.3~keV with low probability (see \fig \ref{pic:DecayScheme}). 

%
%
%
%

\begin{figure}[h]
\centering
\includegraphics[width=0.75\textwidth]{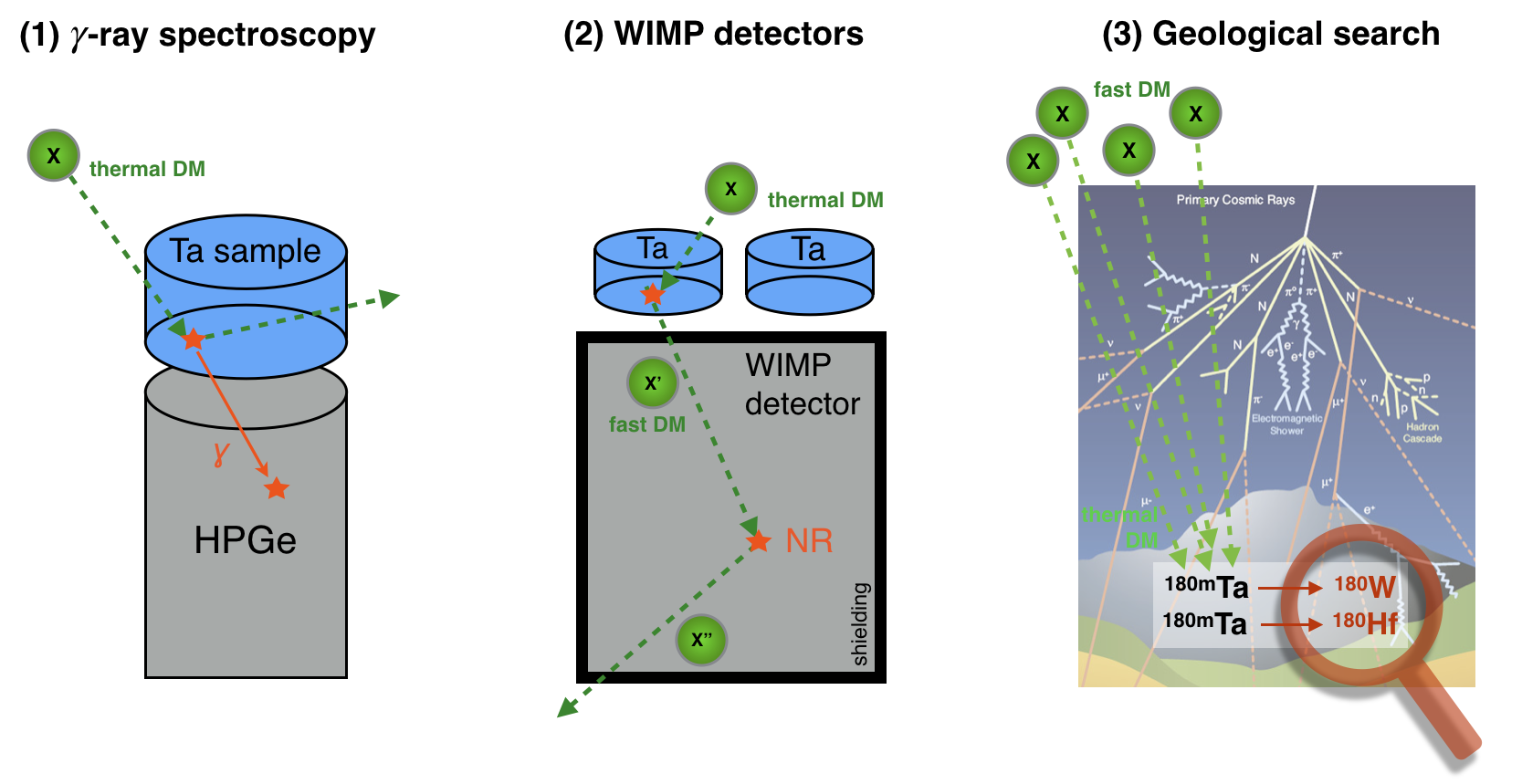}%
\caption{\label{pic:ConceptOverview} Three different concepts to search for dark matter with \nuc{Ta}{180m}: \gray\ spectroscopy, nuclear recoil WIMP detectors, and geological searches. }
\end{figure}

A first time observation of the \nuc{Ta}{\rm 180m} decay would be very interesting by itself, giving insight into the complex nature of highly forbidden nuclear transitions. 
However, the opportunity to use the unique features of \nuc{Ta}{180m} as a new DM detector is even more compelling. 
At least three types of experiments can be envisioned which are illustrated in \fig \ref{pic:ConceptOverview}. 
(1) The \gray\ signature of a DM induced decay can be observed with e.g.\ HPGe \gray\ spectroscopy. 
(2) The stored energy in the isomer can accelerate DM particle making them visible in standard WIMP detectors when large amounts of tantalum are placed near the active matrtial.
(3) Tantalum minerals in geological deposits have experienced a possible DM exposure for much of earths history. An excess of the decay daughters in such minerals can be a DM signature.

For all three types, the DM de-excitation rate will be overburden dependent for some of the DM models. In case of strongly interacting DM, its accumulated density increases towards the center of the earth.   
For type (1) and (3) the experimental signature can be identical to the Standard Model internal conversion or $\gamma$-decay of \nuc{Ta}{180m}. As long as the signature is not observed, limits will constrain both the Standard Model decay and DM de-excitation. If the decay will be observed in the future, multiple measurements showing a decrease of the \nuc{Ta}{180m} half-life at deeper locations will be the smoking gun.

\section{Search with $\bf \gamma$-ray Spectroscopy}

The search for \nuc{Ta}{180m} decay with \gray\ spectroscopy has a long history going back to 1958. The search in Ref.\ \cite{Lehnert:2017} gives the most stringent constraints for the $\beta$-decay and electron capture modes. The same datasets were used in Ref.\ \cite{Lehnert:2020} to constrain the $\gamma$-decay and internal conversion modes. Therein, also the first DM de-excitation constrains were presented.

\begin{figure}[h]
\centering
\includegraphics[width=0.99\textwidth]{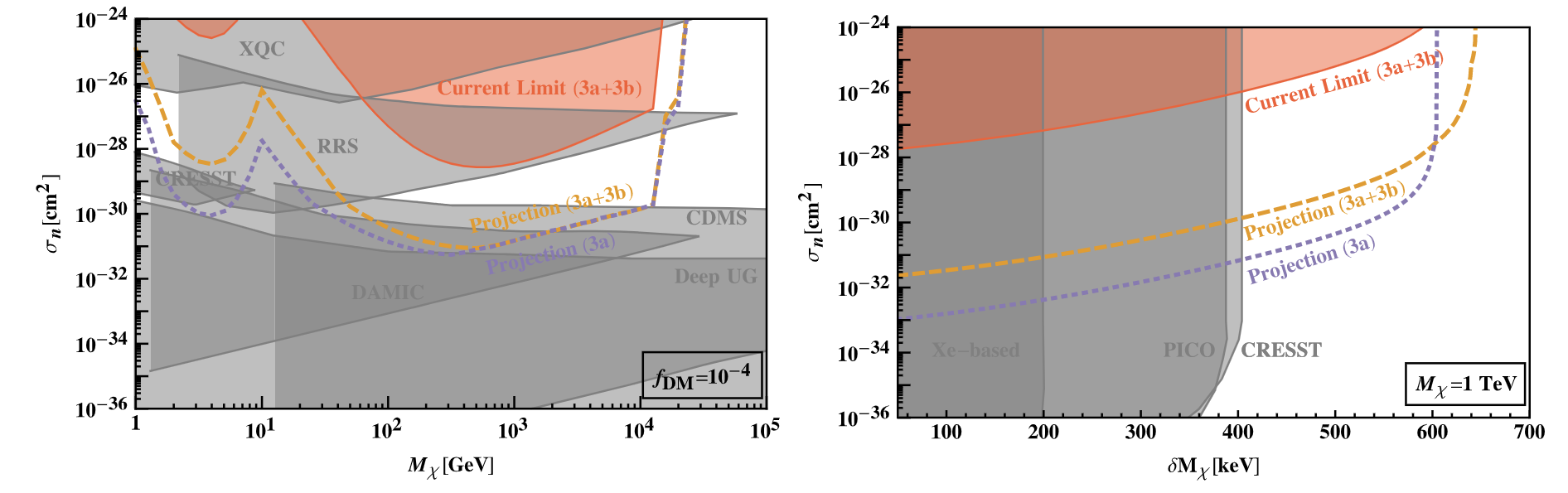}%
\caption{\label{pic:DMExclusion} Exclusion plots for two DM scenarios. Left: Strongly interacting dark matter if it is a fraction of \baseTsolo{-4} of the galactic DM density. Right: Inelastic DM with $\delta M_{\chi}$ excitation energy. Red regions are newly excluded, gray regions were previously excluded by other experiments, and yellow and violet contours are projections for future experiments. From \cite{Lehnert:2020}.}
\end{figure}

The data was taken at the HADES underground lab at a depth of 223~m. For strongly interacting DM, thermalizing in the atmosphere and overburden, this creates a local over-density of about \baseTsolo{8} compared to the local galactic density of 0.3~GeV/cm$^3$.
The half-life limit of \baseT{1.3}{14}~yr was converted into exclusion plots for two DM scenarios shown in \fig \ref{pic:DMExclusion}.
On the left is the case of strongly interacting DM as a fraction of \baseTsolo{-4} of the overall galactic DM density. The parameter space is the per-nucleon cross section vs the DM mass. The gray areas are excluded by previous experiments. The red area is newly excluded. The yellow and violet contours are sensitivity estimates for realistic future experiments. Larger fractions of strongly interacting DM are better constrained by previous experiments (plots can be found in Ref.\ \cite{Lehnert:2020}). Nonetheless, the \nuc{Ta}{180m} method offers an entirely different way of probing this parameter space, strengthening existing limits.
The plot on the right shows the parameter space for composite inelastic DM. If DM scatters require the excitation of the DM from $M_{\chi}$ to $M'_{\chi}$ with $M'_{\chi} - M_{\chi} = \delta M_{\chi}$, the material can provide extra energy for $\delta M_{\chi}$. \nuc{Ta}{180m} provides up to 77~keV and can probe a larger parameter space in $\delta M_{\chi}$ compared to conventional detectors such as PICO and CRESST which only rely on the kinetic energy of the DM for excitation.

Experimentally, only the 103.5~keV \gline\ from the $\beta$-decay branch was accessible. The 93.3~keV \gline\ from the EC branch has natural decay chain background \glines\ from \nuc{Th}{234} at similar energies and could not be used. Future experiments will need to be radiopure, tuned for high detection efficiency of low energy \grays, and ideally use tantalum enriched in \nuc{Ta}{180} reducing self-absorption in the sample.

\section{Search with existing WIMP Detectors}

Existing WIMP detectors could be used to search for exotic DM when surrounded by a large mass of tantalum. Without tantalum, thermalized DM has insufficient kinetic energy to create a detectible nuclear recoil. With tantalum in the vicinity, DM particles could gain kinetic energy from interacting with \nuc{Ta}{180m}, sufficient enough to create detectable nuclear recoils. This is essentially a source on-off experiment for WIMP detectors.

In practice, the expected recoil signature depends heavily on the DM mass, its cross-section, and experimental setup. The largest energy transfer from \nuc{Ta}{180m} to DM is expected for equal masses, i.e.\ at 180~GeV DM mass. In order to not infer with the background in WIMP detectors, the tantalum should be places outside of the shielding. Shielding is often designed to stop neutrons, containing low A material such as hydrogen or carbon. In such a case, DM with 180~GeV mass - even if strongly interacting - might penetrate existing shielding.
Realistic predictions are very dependent on the existing experimental setup and should be simulated with particle propagation, taking into account various DM properties. We thus do not discuss this concept further. 
However, we note that radiopurity requirements of the tantalum might be low outside of the experimental shielding and the cost for large amounts of commercial tantalum might approach the market prize of about 150 USD/kg, allowing potentially large masses.

\section{Geological Search}

\begin{figure}[h]
\centering
\includegraphics[width=0.45\textwidth]{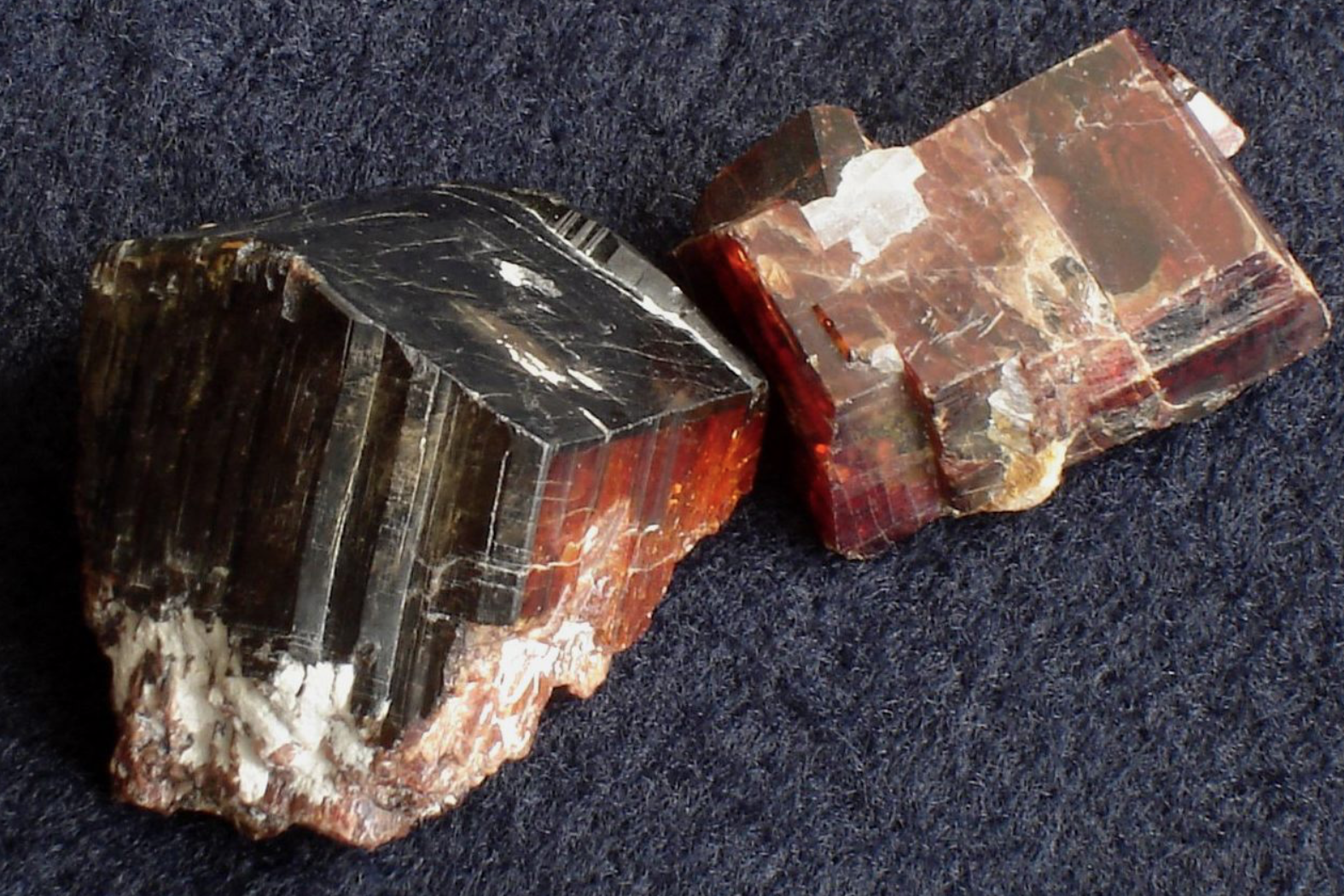}
\includegraphics[width=0.4\textwidth]{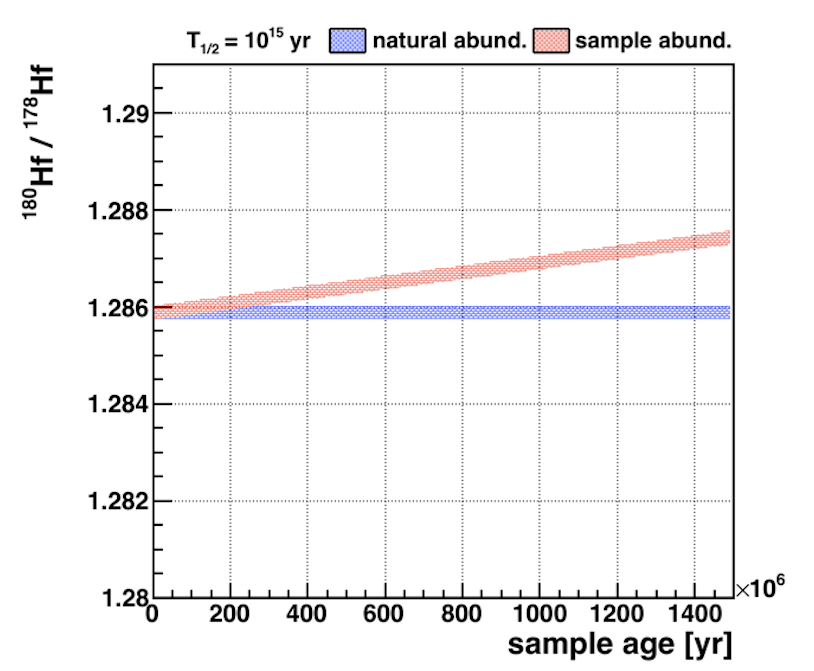}

\caption{\label{pic:tantalite} Left: Tantalite crystal [wiki commons]. Right: Experimental signature of isotopic ratios with and without sample for a given half-life of \baseTsolo{15}~yr and reasonable measurement uncertainties as described in the text. This is not a measurement. }

\end{figure}

In nature, tantalum occurs in tantalite minerals (\fig \ref{pic:tantalite} left). In tantalite, \nuc{Ta}{\rm 180m} has time to decay and accumulate its decay products over geological time scales. This increases the``lifetime of the experiment" by 6 to 9 orders of magnitude compared to typical lab experiments \cite{Mackay:2014by}. 
However, this comes with significant systematic uncertainties on all input parameters such as the geological history of the minerals, chemical extraction efficiencies, and trace concentration of the daughters isotopes in tantalite. Furthermore, giving up on direct detection techniques requires good control of background processes which could lead to the same decay products of \nuc{W}{180} and \nuc{Hf}{180}.

In the past, geochemical experiments have been successfully used for double beta decay searches in Te, Se, and Ba isotopes. In these cases, the daughter isotopes are noble gasses (Xe and Kr) with typically very low trace concentrations in earth minerals. This makes those experiments very sensitive, correctly measuring long half-lives around \baseTsolo{20-24}~yr \cite{GeoChemTe,GeoChemSe,GeoChemBa} of which many were later confirmed by direct detection experiments. However, the search for decay daughters could not distinguish between the standard and the neutrinoless double beta decay modes and they were eventually given up.

For \nuc{Ta}{180m}, the daughters of \nuc{W}{180} and \nuc{Hf}{180} might be present in earth's minerals in much higher quantities  than noble gasses \cite{traceAbun}, likely limiting the sensitivity to much shorter half-lives. However, for \nuc{Ta}{180m}, geochemical experiments include information on the decay mode via the ratio of two daughter isotopes making this approach much more powerful. For the \nuc{Ta}{180} decay a \nuc{Hf}{180} / \nuc{W}{180} ratio of 5.67 is expected whereas for the direct \nuc{Ta}{180m} decay a ratio of 4000 is expected based on the inverse theoretical half-life predictions in Ref.\ \cite{Ejiri:2017js}.  

The isotopic abundances of \nuc{W}{180} and \nuc{Hf}{180} would be measured w.r.t.\ reference isotopes (e.g. \nuc{W}{184} and \nuc{Hf}{178}), not affected by \nuc{Ta}{180m} decay. The experimental signature is the higher relative abundance of the daughters in tantalite w.r.t.\ the natural relative abundance in literature or dedicated reference measurements in local minerals without tantalum. 
\fig \ref{pic:tantalite} (right) shows the abundance ratio of \nuc{Hf}{180} to \nuc{Hf}{178}. The natural abundance ratio (blue) is stable at 1.286 \cite{isotopicAbun2016}. In typical tantalite, the ratio (red) would increase with the sample age, in this case assuming a \nuc{Ta}{180m} half-life of \baseTsolo{15}~yr. The width of the lines show the 1$\sigma$ uncertainty from mass spectroscopy (MS) in the literature. Newer MS techniques might reduce this uncertainty significantly.
This example is based on realistic inputs, but many parameters are still unknown. Nevertheless, with a 1 billion year old sample, a significant decay signal could be observed for a \baseTsolo{15}~yr \nuc{Ta}{180m} half-life. This is a factor of 10 better compared to current direct detection limits. 
A different measured half-life in minerals of different depth would be a smoking gun for DM induced de-excitation.\\


\end{document}